\documentclass[letter,traditabstract]{aa}
\usepackage{natbib}
\usepackage{lscape}
\usepackage[section]{placeins}
\usepackage{wasysym}
\usepackage{graphics}
\usepackage{graphicx}
\usepackage{txfonts}
\usepackage{color}
\usepackage{times}
\usepackage{mathtools}
\usepackage{changes}
\usepackage{hyperref}
\usepackage{ulem}     
\hypersetup{
    pdftoolbar=true,            
    pdfmenubar=true,            
    pdffitwindow=false,         
    pdfstartview={FitH},        
    pdftitle={title},                   
    pdfauthor={Author},         
    pdfsubject={Subject},       
    pdfcreator={Author},        
    pdfproducer={Author},       
    pdfkeywords={AGN} {Stellar Content} , 
    pdfnewwindow=true,  
    colorlinks=true,            
    linkcolor=cyan,             
    citecolor=cyan,             
    filecolor=cyan,                     
    urlcolor=cyan               
}

\newcommand{\spl}{\texttt{spl}}
\newcommand{\mpl}{\texttt{mpl}}

\newcommand{\Starlight}{{\sc Starlight}}
\newcommand{\Rebetiko}{{\sc Rebetiko}}

\begin{document}

%
\titlerunning{AGN detection threshold in Lyman-continuum-leaking ETGs}  
\authorrunning{L. S. M. Cardoso, J. M. Gomes \& P. Papaderos}

\title{Semi-empirical AGN detection threshold in spectral synthesis studies of Lyman-continuum-leaking early-type galaxies}  

\author{
Leandro S. M. Cardoso\inst{\ref{adress1},\ref{adress2}}
\and Jean Michel Gomes \inst{\ref{adress1}}
\and Polychronis Papaderos\inst{\ref{adress1}}
}

\institute{
Instituto de Astrof\' isica e Ci\^ encias do Espa\c co, Universidade do Porto, CAUP, Rua das Estrelas, PT4150-762 Porto, Portugal\label{adress1}\\
\email{\href{mailto:Leandro.Cardoso@astro.up.pt}{Leandro.Cardoso@astro.up.pt}}
\and
Departamento de F\' isica e Astronomia, Faculdade de Ci\^encias, Universidade do Porto, Rua do Campo Alegre 687, PT4169-007 Porto, Portugal\label{adress2}
}
\date{Received ?? / Accepted ??}

\abstract{Various lines of evidence suggest that the cores of a large portion of early-type galaxies (ETGs) are virtually evacuated of warm ionised gas. This implies that the Lyman-continuum (LyC) radiation produced by an assumed active galactic nucleus (AGN) can escape from the nuclei of these systems without being locally reprocessed into nebular emission, which would prevent their reliable spectroscopic classification as Seyfert galaxies with standard diagnostic emission-line ratios. The spectral energy distribution (SED) of these ETGs would then lack nebular emission and be essentially composed of an old stellar component and the featureless power-law (PL) continuum from the AGN. A question that arises in this context is whether the AGN component can be detected with current spectral population synthesis in the optical, specifically, whether these techniques effectively place an AGN detection threshold in LyC-leaking galaxies.
To quantitatively address this question, we took a combined approach that involves spectral fitting with \Starlight\ of synthetic SEDs composed of stellar emission that characterises a 10 Gyr old ETG and an AGN power-law component that contributes a fraction  $0\leq x_{\mathrm{AGN}} < 1$ of the monochromatic luminosity at $\lambda_0=$ 4020 \AA.  In addition to a set of fits for PL distributions $F_{\nu} \propto \nu^{-\alpha}$ with the canonical $\alpha=1.5$, we used a base of  multiple PLs with $0.5 \leq \alpha \leq 2$ for a grid of synthetic SEDs with a signal-to-noise ratio of 5--$10^3$. 
Our analysis indicates an effective AGN detection threshold at $x_{\mathrm{AGN}}\simeq 0.26$, which suggests that a considerable fraction of ETGs hosting significant accretion-powered nuclear activity may be missing in the AGN demographics.  
}


\maketitle
\section{Introduction}\label{Sec:Introduction}

        The build-up of super-massive black holes (SMBHs) and co-evolution with their galaxy hosts is a central question in extragalactic astronomy.  In this regard, comprehensive multi-wavelength studies of matter accretion onto SMBHs and the associated active galactic nuclei (AGN) phenomenon are  fundamental for elucidating the interplay between SMBH growth and galaxy assembly history.  Accretion-powered activity in galaxies is mostly being addressed in the optical through emission-line ratio diagnostics,  which allow distinguishing between, for instance, gas photoionisation by an AGN and OB stars (e.g. \citealt{Baldwin_Phillips_Terlevich_1981, Veilleux_Osterbrock_1987}).
        The diagnostic power of such methods presumably vanishes, however, when galactic nuclei are mostly evacuated of gas (e.g. permeated by hot and tenuous plasma), thereby allowing the bulk of the Lyman continuum (LyC) output from an AGN to escape without being locally reprocessed as nebular emission. In those cases the optical spectral energy distribution (SED) is expected to essentially consist of stellar emission and a featureless AGN continuum. In the limiting case of a dominant AGN and high LyC-photon escape fraction, the optical SED of a galaxy is therefore expected
to entirely lack both  nebular (continuum and line) emission and stellar absorption features and superficially somewhat resemble that of a BL Lac (e.g. \citealt{Oke_Gunn_1974, Marcha_etal_1996, Kugler_etal_2014}).

        The situation of an AGN leaking LyC radiation envisaged here may apply to many early-type galaxy (ETG) nuclei. \cite{Papaderos_etal_2013} have first estimated the LyC escape fraction from ETG nuclei to be 70--95\% (see also \citealt{Gomes_etal_2016}) and pointed out that LyC photon escape may constitute a key element in understanding why many of these systems, despite evidence of a strong AGN energy source in radio or X-ray wavelengths, show only weak optical emission-lines typical of low-ionization nuclear emission-line regions (LINERs, \citealt{Heckman_1980}). Several questions naturally arise from such considerations. For instance, to which extent can an AGN be recovered from the optical SED of such LyC-leaking ETGs using currently available spectral fitting techniques? More specifically, do these techniques effectively place a detection threshold on the featureless AGN continuum, and are they capable of recovering its main characteristics (e.g. luminosity contribution, spectral slope) simultaneously with those of the stellar component in the host galaxy?

        Spectral population synthesis (SPS, also referred to as the inverse or semi-empirical approach; \citealt{Faber_1972}) has been applied over the past decades to optical Seyfert galaxy spectra with the goal of evaluating the AGN power-law (PL) featureless continuum and how this might affect the retrieved star formation history (e.g. \citealt{Koski_1978, Schmitt_StorchiBergmann_CidFernandes_1999, CidFernandes_etal_2004, Benitez_etal_2013}). However, none of these works have attempted to quantitatively infer an AGN detectability threshold for LyC-leaking ETGs, which is the goal of this study. 

\section{Method and results}\label{Sec:Methodology_Results}

        The approach taken here involves computing and subsequently modelling synthetic SEDs with an SPS code. For the first task we computed a synthetic optical spectrum with our evolutionary synthesis code \Rebetiko\ (Papaderos \& Gomes, in prep.)  that
represents an ETG that has formed monolithically 10 Gyr ago. For this, we adopted simple stellar population (SSP) spectra of solar metallicity ($Z_{\odot}=0.02$) from \citet{Bruzual_Charlot_2003}, with a \citet{Chabrier_2003} initial mass function (IMF) between 0.1 and 100 $M_{\odot}$ and Padova evolutionary tracks (\citealt{Alongi_etal_1993, Bressan_etal_1993, Fagotto_etal_1994a, Fagotto_etal_1994b, Fagotto_etal_1994c, Girardi_etal_1996}), and zero intrinsic extinction in the V band ($A_V=0$ mag). The featureless AGN continuum, which was added to the stellar SED, was parameterised by a PL defined as $F_{\nu} \propto \nu^{-\alpha}$ (e.g. \citealt{Oke_Neugebauer_Becklin_1970, OConnell_1976, Koski_1978}) with the canonical spectral index $\alpha=1.5$ for Seyfert galaxies (e.g. \citealt{Ferland_Netzer_1983, Schmitt_StorchiBergmann_CidFernandes_1999, Kauffmann_etal_2003c, CidFernandes_etal_2004}). Afterwards, we attempted to recover the AGN characteristics ($\alpha$, light fraction) by fitting the synthetic SEDs with the most recent
publicly available version of the SPS code \Starlight\ (\citealt{CidFernandes_etal_2005}, public distribution v04).           
        
\begin{figure} [t!]
\begin{center}
\includegraphics[width=0.48\textwidth]{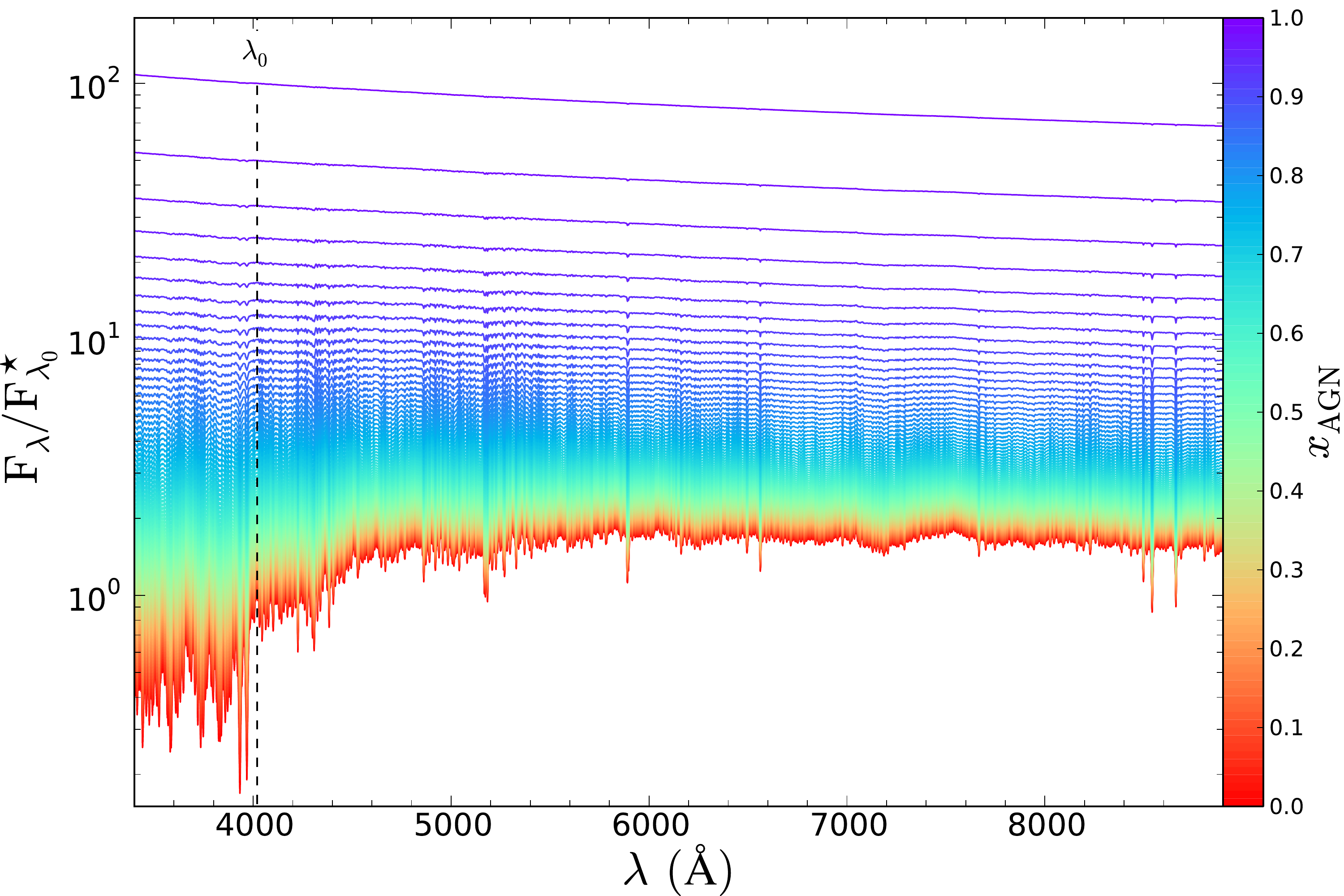}
\caption{ Synthetic galaxy spectra resulting from the combination of a 10 Gyr old solar-metallicity stellar population  with a power-law AGN continuum defined as $F_{\nu}\propto \nu ^{-1.5}$. The colour coding represents the AGN fractional flux contribution $x_{\mathrm{AGN}}$ at $\lambda_0=$ 4020 \AA. }
\label{Fig:Spectral_variations_in_xagn}
\end{center}
\end{figure}

        The continuum flux $F_{\lambda}$ at the normalisation wavelength $\lambda_0$ resulting from the superposition of the AGN PL $F^{\mathrm{AGN}}_{\lambda}$ to the stellar SED $F^{\star}_{\lambda}$ can be written as
        \begin{equation}\label{Eq:Linear_Combination_Modified}
        F_{\lambda} = F^{\star}_{\lambda} + \frac{x_{\mathrm{AGN}}}{x_{\star}}  F^{\mathrm{AGN}}_{\lambda}
                                = F^{\star}_{\lambda} + \frac{x_{\mathrm{AGN}}}{x_{\star}}  
                                  F^{\star}_{\lambda_0} \left( \frac{\lambda}{\lambda_0} \right)^{\alpha-2} ,
                                                                                                                                                                \end{equation}  
        \noindent where $x_{\star}$ and $x_{\mathrm{AGN}}$ are the fractional stellar and AGN flux contributions, respectively, constrained by the normalisation condition $x_{\star}+x_{\mathrm{AGN}} = 1$ at $\lambda_0=4020$ \AA. Equation \ref{Eq:Linear_Combination_Modified} ensures that the stellar physical properties (e.g. flux, mass) computed with \Rebetiko\ are conserved by the addition of the AGN component.
%
\begin{figure}[t!]
\begin{center}
\includegraphics[width=0.47\textwidth]{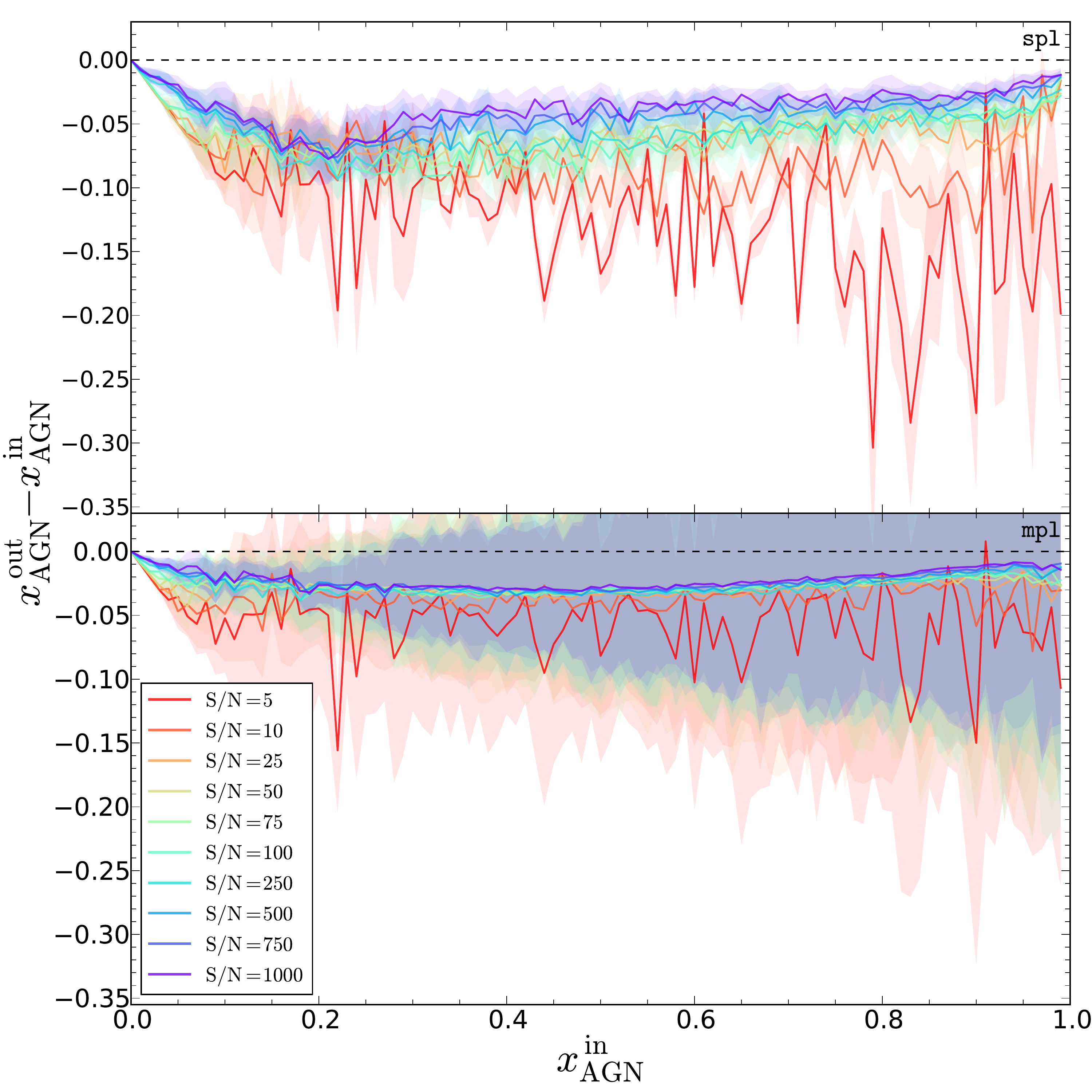}
\caption{Difference between output and input AGN fractional flux contribution $x_{\mathrm{AGN}}^{\mathrm{out}} - x_{\mathrm{AGN}}^{\mathrm{in}}$  at $\lambda_0$ as a function of $x_{\mathrm{AGN}}^{\mathrm{in}}$ for ten S/N values. Results obtained by fitting synthetic spectra with \Starlight\  with a single (\spl) and multiple (\mpl) AGN power-law components in the base library are shown in the upper and lower panels, respectively. The shaded area corresponds to the $\pm 1\sigma$ standard deviation around the mean, as computed from ten \Starlight\ fits to each synthetic spectrum. 
}
\label{Fig:xagn_diff_as_a_function_of_xagn}
\end{center}
\end{figure}

        In this work we adopted a PL index $\alpha=1.5$ and a range for the AGN relative contribution of $x_{\mathrm{AGN}} \in [0,1[$ in steps of 0.01 to create synthetic spectra, as shown in Fig. \ref{Fig:Spectral_variations_in_xagn}. We note that SEDs with $x_{\mathrm{AGN}}\geq 0.5$ are only included for the sake of completeness,  given that already an $x_{\mathrm{AGN}}\ga 0.4$ corresponds to conspicuously featureless spectra that would normally  be excluded from SPS modelling upon visual inspection.  Finally, the set of synthetic SEDs comprises ten signal-to-noise (S/N) ratios between 5 and $10^3$ at $\lambda_0$ in order to study how the AGN recovery depends on the quality of input spectra.
        
        The synthetic SEDs were then fitted with \Starlight\ using a base with SSPs from \cite{Bruzual_Charlot_2003} for 25 ages from 1 Myr to 13 Gyr and 4 metallicities ($Z = 0.2$, 0.4, 1 and 2.5 $Z_{\odot}$) with a \cite{Chabrier_2003} IMF and Padova evolutionary tracks. Additionally, the base contains a single PL with $\alpha=1.5$, as the one embedded in the input SED. The spectral fitting was performed between 3400 and 8900 \AA\ following common practice in studies of local galaxy samples from SDSS (e.g. \citealt{Asari_etal_2007, Ribeiro_etal_2016}), with $A_V$ kept as a free parameter between -1 and 4 mag. Moreover, Starlight was applied ten times for each spectrum assuming different initial guesses in the parameter space (i.e. seed numbers in the fitting procedure) to better evaluate formal uncertainties. 

        A second set of fits was computed using seven PLs in the base with an $\alpha$ between 0.5 and 2 in steps of 0.25, within the range of values commonly adopted both in spectral synthesis (e.g. \citealt{Goerdt_Kollatschny_1998, Schmitt_StorchiBergmann_CidFernandes_1999, Kauffmann_etal_2003c, CidFernandes_etal_2004, Moultaka_2005, Benitez_etal_2013}) and photoionisation models (e.g. \citealt{Ferland_Netzer_1983, Stasinska_1984a,Stasinska_1984b, Mathews_Ferland_1987, Veilleux_Osterbrock_1987}). The motivation behind this experiment was to check whether \Starlight\ can choose the PL with the correct slope, a test which is of relevance to its unsupervised application to large spectroscopic data sets without a priori knowledge of $\alpha$. The allowance for multiple PLs of different $\alpha$ obviously bears the risk that an eventual inclusion of more than one PL in the best-fitting solution would lead to a non-PL distribution, which in turn might affect the properties inferred for the stellar component (e.g. $A_V$).  These two series of fits are referred to in the following as \spl\ (single PL) and \mpl\ (multiple PLs), respectively.   
        
        For both \spl\ and \mpl, it is important to bear in mind that the results presented below correspond to a best-case scenario where the ingredients used to compute the input SEDs (i.e. SSPs and PLs) are also included in the base that is subsequently used for spectral fitting. This, and the fact that synthetic SEDs are composed of only a single-age stellar component, instead of the rather general case of a composite stellar population, further simplifies the framework of this study and facilitates the recovery of the AGN characteristics.  Thus, the following analysis probably yields a generous estimate on the true capability of \Starlight\ for recovering an AGN hosted within a LyC-leaking ETG.
        
\begin{figure} [t!]
\begin{center}
\includegraphics[width=0.47\textwidth]{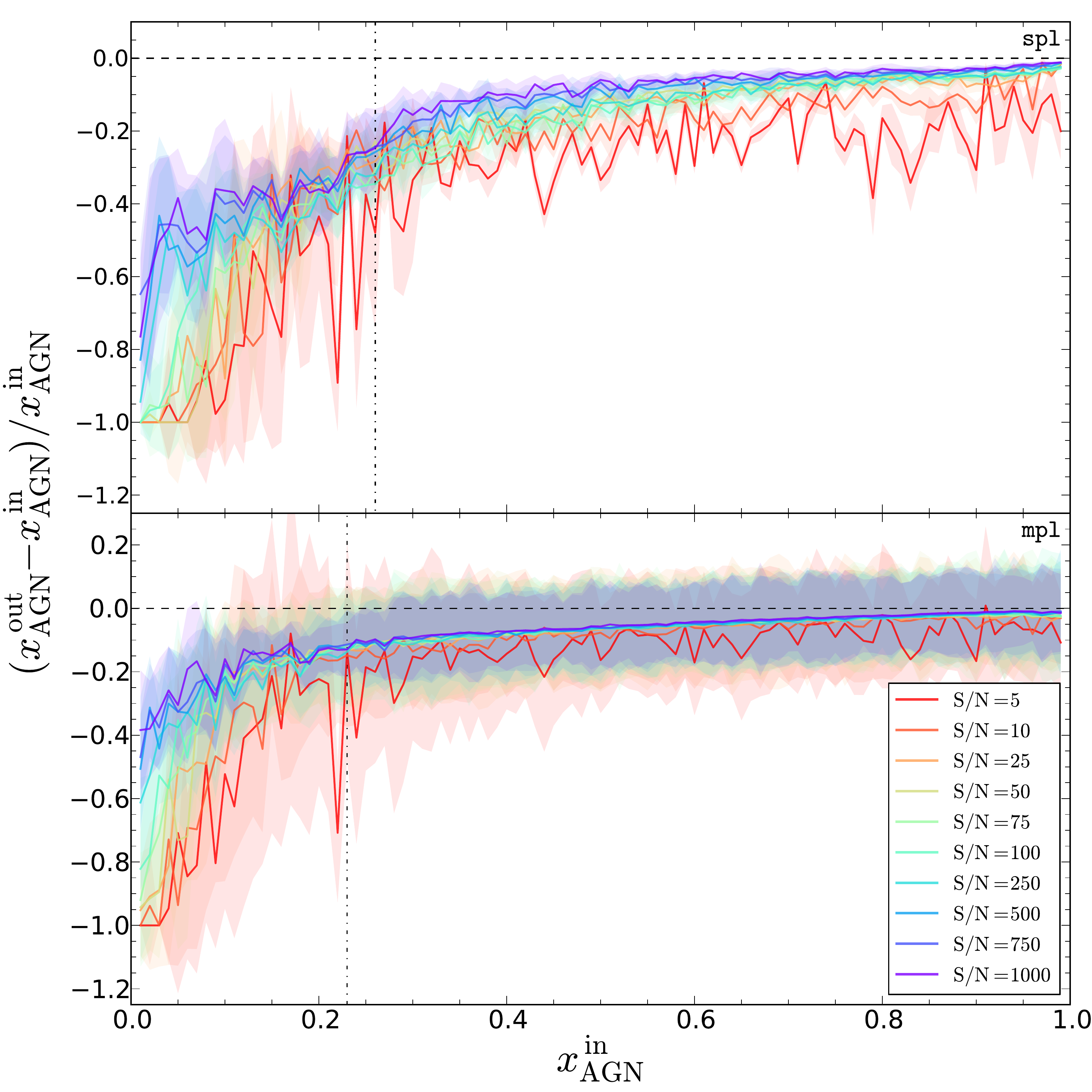}
\caption{Relative error of the fractional flux contribution of the AGN $({x_{\mathrm{AGN}}^{\mathrm{out}} - x_{\mathrm{AGN}}^{\mathrm{in}}}) / {x_{\mathrm{AGN}}^{\mathrm{in}}}$ as a function of $x_{\mathrm{AGN}}^{\mathrm{in}}$ for \spl\ and \mpl\ fits (upper and lower panels, respectively). The vertical dash-dotted lines mark the AGN detection threshold inferred from the \cite{Gelman_Rubin_1992} convergence test (see discussion for details). }
\label{Fig:xagn_fractional_error_as_a_function_of_xagn}
\end{center}
\end{figure}

        The first step in our analysis is to examine to which extent \Starlight\ can recover the AGN contribution. Figure \ref{Fig:xagn_diff_as_a_function_of_xagn} shows the difference between the output and input luminosity contribution $x_{\mathrm{AGN}}^{\mathrm{out}} - x_{\mathrm{AGN}}^{\mathrm{in}}$ of the AGN as a function of $x_{\mathrm{AGN}}^{\mathrm{in}}$ for ten S/N values (see legend for details). The accuracy on the estimation of $x_{\mathrm{AGN}}$ slightly increases with increasing AGN contribution, despite several local deviations from the true value, especially for lower-S/N ($\la25$) spectra. A salient feature in both panels is that \Starlight\  on average underestimates the fractional contribution of the PL by $\sim$6.2\% ($\sim$3.0\%)  for \spl\ (\mpl) models, and that this offset, even though subtle, is apparent over the entire $x_{\mathrm{AGN}}$ range. 

\begin{figure} [t!]
\begin{center}
\includegraphics[width=0.47\textwidth]{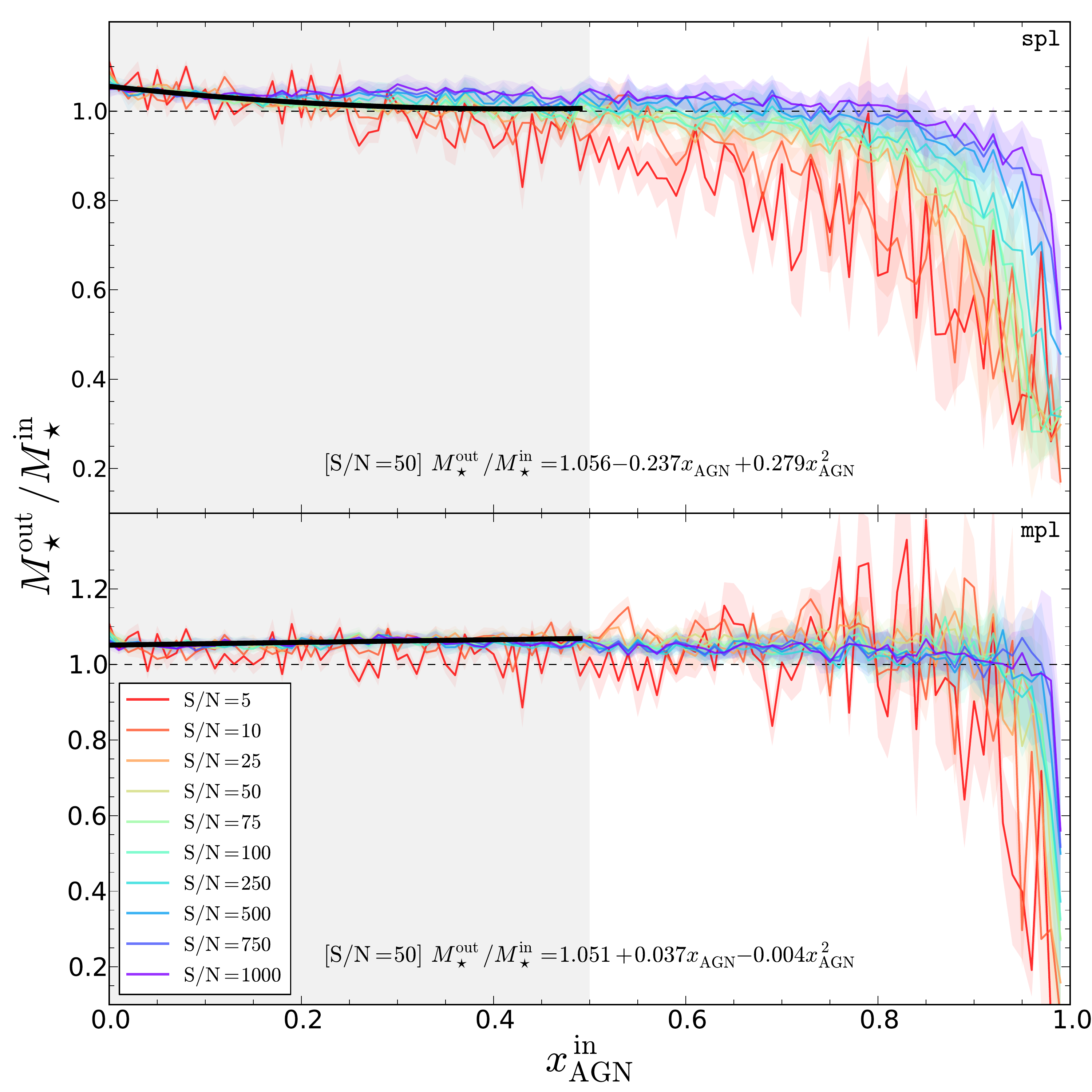}
\caption{Ratio between output and input total stellar mass $M_{\star}$ as a function of $x_{\mathrm{AGN}}^{\rm in}$ 
for \spl\ and \mpl\ fits (upper and lower panels, respectively). Thick black lines represent second-order polynomial fits  for $x_{\mathrm{AGN}}^{\text{in}} \leq 0.5$ and a  S/N=50.}
\label{Fig:M_ratio_as_a_function_of_xagn}
\end{center}
\end{figure}

        This bias is best visible in Fig. \ref{Fig:xagn_fractional_error_as_a_function_of_xagn}, which shows the relative deviation  $({x_{\mathrm{AGN}}^{\mathrm{out}} - x_{\mathrm{AGN}}^{\mathrm{in}}}) / {x_{\mathrm{AGN}}^{\mathrm{in}}}$. The AGN contribution is systematically underestimated, reaching $\sim$35.4\% and $\sim$14.9\% at $x_{\mathrm{AGN}}=0.2$ for \spl\ and \mpl, respectively. Interestingly, the deviation is on average by $\sim$10.7\% higher for \spl\ fits, although these employ the single PL component embedded in the input SED. 
        Assuming that the lines inferred for each S/N in Fig. \ref{Fig:xagn_fractional_error_as_a_function_of_xagn} are analogues of independent chains in a random-walk Monte Carlo simulation, we can estimate an AGN detection threshold from the \citet{Gelman_Rubin_1992} convergence criterion.   This yields a variance ratio $\hat{R}<1.1$ for $x_{\mathrm{AGN}} \geq 0.26$ and 0.23 (\textup{\textit{dash-dotted lines}}) for \spl\ and \mpl\ fits, respectively. These values reflect the empirical insight  from Fig. \ref{Fig:Spectral_variations_in_xagn}  that traces of a diluting AGN continuum are barely recognisable by eye  for a $x_{\mathrm{AGN}}\la 0.2$.  Automated application of \Starlight\ on large spectroscopic data sets (e.g. SDSS, 6dF, GAMA) would thus underestimate or entirely miss the AGN contribution, potentially leading to the erroneous classification of LyC-leaking AGNs as lineless \emph{passive} galaxies, in the notation by \citet{CidFernandes11}.  It is also noteworthy that the stellar mass $M_{\star}$ is recovered in either case at a level better than $\sim$10\% for ${x_{\mathrm{AGN}}^{\mathrm{in}}} \la 0.5$ (shaded area in Fig. \ref{Fig:M_ratio_as_a_function_of_xagn}).

\begin{figure} [t!]
\begin{center}
\includegraphics[width=0.47\textwidth]{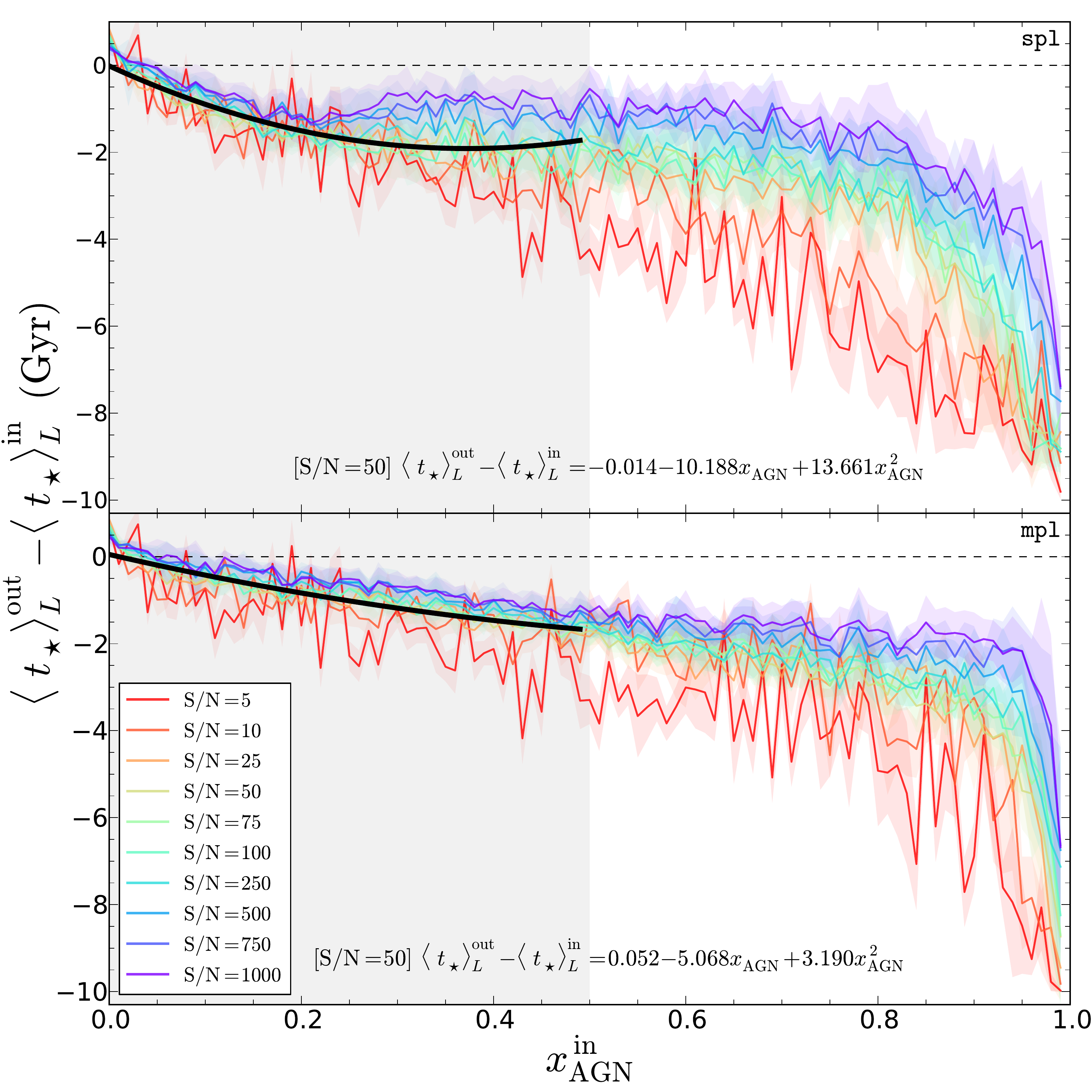}
\caption{Difference between the output and input luminosity-weighted mean stellar age $\langle t_{\star} \rangle_L$ as a function of $x_{\mathrm{AGN}}^{\rm in}$ for \spl\ and \mpl\ models  (upper and lower panels, respectively). Thick black lines have the same meaning as in Fig.~\ref{Fig:M_ratio_as_a_function_of_xagn}. }
\label{Fig:age_L_difference_as_a_function_of_xagn}
\end{center}
\end{figure}

\begin{figure}[t!]
\begin{center}
\includegraphics[width=0.47\textwidth]{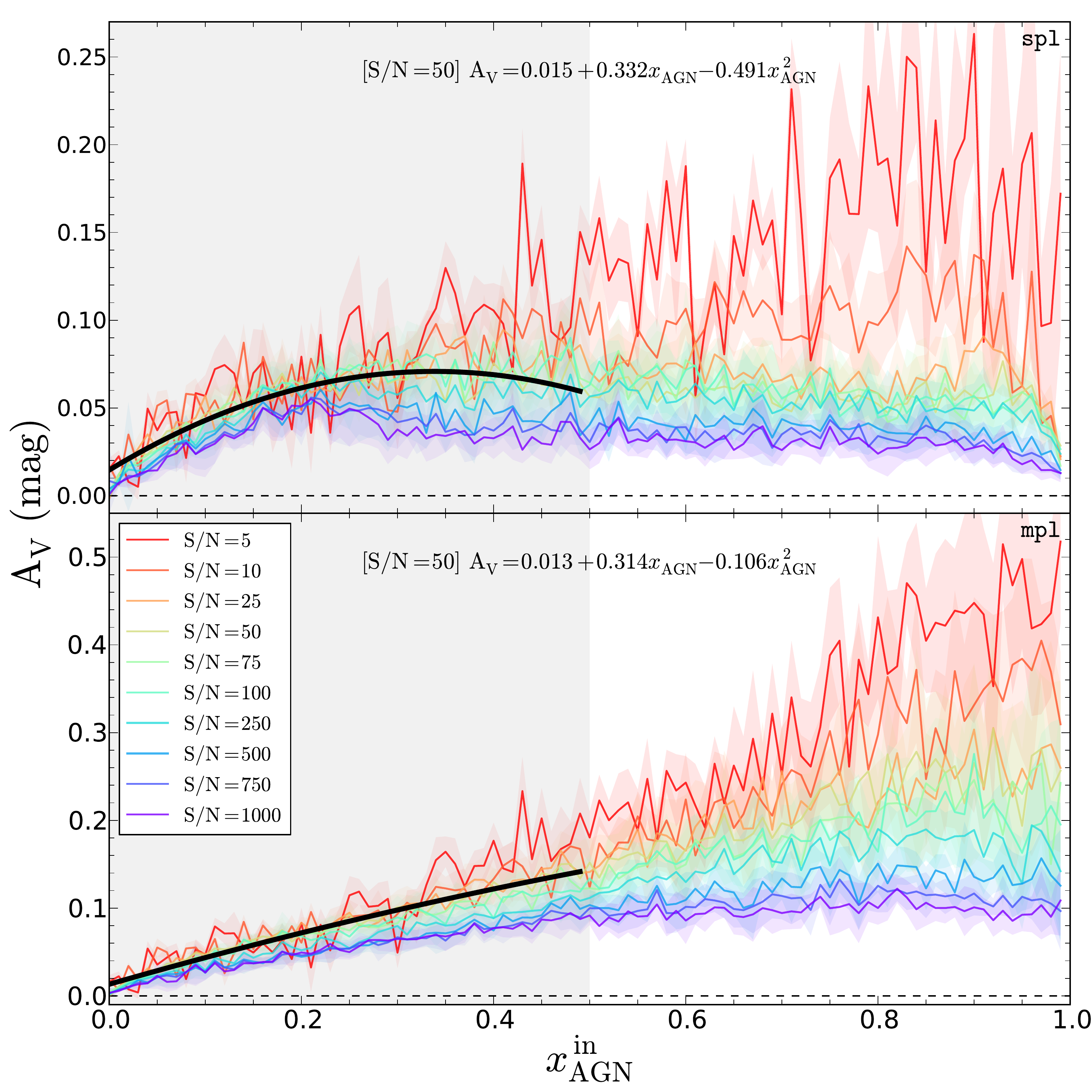}
\caption{Inferred intrinsic extinction $A_V$ (mag) in the $V$ band as a function of $x_{\mathrm{AGN}}^{\rm in}$ for \spl\ and \mpl\ fits (upper and lower panels, respectively). The input SEDs are computed for an $A_V=0$.  Thick black lines have the same meaning as in Fig.~\ref{Fig:M_ratio_as_a_function_of_xagn}.}
\label{Fig:extinction_as_a_function_of_xagn}
\end{center}
\end{figure}

        One question that arises from the foregoing is whether the underestimation of the AGN contribution despite almost unbiased $M_{\star}$ estimates could come at the price of biases in other quantities of relevance, such as  the luminosity-weighted mean stellar age $\langle t_{\star} \rangle_L$ and the intrinsic extinction $A_V$.  This is indeed the case judging from Figs. \ref{Fig:age_L_difference_as_a_function_of_xagn} and \ref{Fig:extinction_as_a_function_of_xagn}, which reveal a trend for a S/N-dependent underestimation of the age and overestimation of $A_V$ over the considered $x_{\text{AGN}}$ range. Specifically, the luminosity-weighted mean stellar age computed with \Starlight\ is underestimated by $\sim$1.7 and 0.8 Gyr at $x_{\text{AGN}}=0.26$ and 0.23 for \spl\ and \mpl, respectively. 
At the AGN detection threshold this bias is apparently partly compensated for by an artificial reddening by $A_V \sim 0.07$ mag. Indirect evidence of a coupling between extinction and age at the expense of $x_{\text{AGN}}$ comes from the fact that the latter is accurately retrieved by \Starlight\ when $A_V$ is fixed to 0 mag.

\newpage
\section{Summary and conclusions}\label{Sec:Summary}

        The aim of this study was to examine whether there is an AGN detectability threshold in optical spectral synthesis studies of LyC-leaking early-type galaxies (ETGs) hosting accretion-powered nuclear activity. To address this question, we took a combined approach that involved spectral fitting with \Starlight\ of synthetic SEDs that are composed of an instantaneously formed 10 Gyr old stellar component \emph{plus} an AGN power-law component providing a fraction $0\leq x_{\mathrm{AGN}} < 1$ of the monochromatic luminosity at $\lambda_0=$ 4020 \AA.  For this set of models, we find that \Starlight\ recovers the stellar mass to within $\sim$10\%, but systematically overestimates the intrinsic extinction and underestimates the luminosity-weighted stellar age and $x_{\mathrm{AGN}}$. 
        Our analysis indicates an effective AGN detection threshold at $x_{\mathrm{AGN}}\simeq 0.26$, which suggests that a considerable fraction of ETGs hosting significant accretion-powered nuclear activity may be missing in the AGN demographics.

\begin{acknowledgements}
We thank the anonymous referee for valuable comments and suggestions. LSMC acknowledges support by Funda\c{c}\~ao para a Ci\^encia e a Tecnologia (FCT) (Ref. UID/FIS/04434/2013) through national funds and by FEDER funding through the program Programa Operacional de Factores de Competitividade (COMPETE) 2020 (Ref. POCI-01-0145- FEDER-007672). JMG acknowledges support by FCT through the Fellowship SFRH/BPD/66958/2009 and POPH/FSE (EC) by FEDER funding through the program COMPETE. PP is supported by FCT through the Investigador FCT Contract No. IF/01220/2013 and POPH/ FSE (EC) by FEDER funding through the program COMPETE. LSMC, JMG \& PP acknowledge support by FCT under project FCOMP-01-0124-FEDER-029170 (Ref. FCT PTDC/FIS-AST/3214/2012), funded by FCT-MEC (PIDDAC) and FEDER (COMPETE). LSMC, JMG \& PP also acknowledge the exchange programme "Study of Emission-Line Galaxies with Integral-Field Spectroscopy" (SELGIFS, FP7-PEOPLE-2013-IRSES-612701), funded by the EU through the IRSES scheme.
\end{acknowledgements}



\end{document}